# RAPID PULSATIONS IN SUB-THz SOLAR BURSTS

Pierre Kaufmann[1,2], C. Guillermo Giménez de Castro[1], Emilia Correia[1,3], Joaquim E.R. Costa[3], Jean-Pierre Raulin[1] and Adriana Silva Válio[1]

[1] *Centro de Rádio-Astronomia e Astrofísica Mackenzie, Escola de Engenharia, Universidade Presbiteriana Mackenzie, Rua Consolação 896, 01302-907 São Paulo, SP, Brazil*
[2] *Centro de Componentes Semicondutores, Universidade Estadual de Campinas, Cidade Universtária" Zeferino Vaz", 13981-970 Campinas, SP, Brazil*
[3] *Instituto Nacional de Pesquisas Espaciais, Av. dos Astronautas 1758, 12227-010 São José dos Campos, SP, Brazil*

## Abstract

A new solar burst emission spectral component has been found showing sub-THz fluxes increasing with frequency, spectrally separated from the well known microwave component. Rapid pulsations are found present in all events observed at the two frequencies of the solar submillimeter-wave telescope (SST): 212 and 405 GHz. They were studied in greater detail for three solar bursts exhibiting the new THz spectral component. The pulse amplitudes are of about 5-8% of the mean flux throughout the bursts durations, being comparable for both frequencies. Pulsations range from one pulse every few seconds to 8-10 per second. The pulse repetition rates (R) are linearly proportional to the mean burst fluxes (S), following the simple relationship S = k R, suggesting that the pulsations might be the response to discrete flare particle accelerator injections quantized in energy. Although this result is consistent with qualitative trends previously found in the GHz range, the pulse amplitude relative to the mean fluxes at the sub-THz frequencies appear to be nearly ten times smaller than expected from the extrapolation of the trends found in the GHz range. However there are difficulties to reconcile the nearly simultaneous GHz and THz burst emission spectrally separated components, exhibiting rapid pulsations with considerably larger relative intensities in the GHz range.

## 1. Introduction

Solar flare impulsive emissions in the radio range are usually explained by two distinct processes: radiation at longer metric-decimetric wavelengths exhibit flux densities decreasing with frequency (usually narrowband), attributed to excited plasma emissions higher in the solar corona (Wild & Smerd 1972). Broadband microwaves are attributed to gyro-synchrotron losses in the magnetic field while accelerated electrons move into denser regions to produce hard X- and gamma rays by collision losses (Dulk, Melrose & White 1979; Kundu & Vlahos 1982; Bastian, Benz & Gary 1998). The two mechanisms compose the well known U-shaped radio burst spectra (Castelli 1972; Nita, Gary & Lee 2004). The THz flux vs. frequency increasing spectral component, recently discovered by Kaufmann *et al.* (2004), suggests a "W-shaped" spectrum as illustrated in Figure 1, possibly representative of a third emission mechanism.

Time profiles at various radio wavelengths and X-ray ranges are not always well correlated. Examples of time profiles with delays of several seconds at short microwaves and hard X-rays (> 30 keV) were found (Takakura *et al.* 1983; Lim *et al*. 1992; Kaufmann *et al.* 1983; Costa & Kaufmann 1983; Silva, Wang & Gary 2000). The burst impulsive phase onset may occur first at lower frequencies, although there are also examples with delays in the opposite sense. Fast superimposed time structures at much shorter second to sub-second scales, obtained by observations with higher sensitivity and time resolution, have indicated that they are synchronized by less than 200 ms at short microwaves and at hard X-rays (Cornell *et al.* 1984; Takakura *et al*. 1983; Kaufmann *et al.* 2000).

Broadband solar burst rapid pulsations at cm-mm wavelengths were well characterized (Butz *et al.* 1976; Gaizauskas & Tapping 1980; Kaufmann *et al.* 1980; 1985; Allaart *et al*. 1990, Benz *et al.* 1992; Qin *et al*. 1996; Huang, Qin & Yao 1996; Raulin *et al.* 1998; Nakajima 2000; Altyntsev *et al.* 2000). A puzzling qualitative proportionality between pulse repetition rates and underlying mean fluxes has been found at 10, 22, 30, 44, 48 and 90 GHz (Kaufmann *et al.* 1980; 1985; Correia & Kaufmann 1987; Raulin *et al.* 1998) and at 9.4 and 15 GHz (Qin *et al*. 1996; Huang, Qin & Yao 1996). It might be interpreted as the response to discrete injections quasi-quantized in energy (Kaufmann *et al.* 1980); by wave-particle plasma instabilities saturation time inversely proportional to radiation time (Huang, Qin & Yao 1996) or by modulation of flare kernel emission by magnetic field oscillations driven by Alfvén waves which period decreases for larger temperatures (Qin *et al.* 1996)

The present study addresses the distinctive aspects of rapid pulsations observed at sub-THz solar bursts. They are analyzed for their repetition rates vs. time profiles at sub-THz frequencies, at hard X-rays and γ-rays and on the differences when compared to pulsations in the microwave range of frequencies.

## 2. The sub-THz solar burst emissions

The solar submillimeter telescope (SST) is operated at the El Leoncito Astronomical Complex (CASLEO), San Juan, Argentina Andes. It has four 212 GHz and two 405 GHz radiometers, placed at the focal plane of a radome-enclosed 1.5-m diameter Cassegrain antenna (Kaufmann *et al.* 2008). Three partial overlapping beams at 212 GHz allow the determination of the accurate burst spatial position and the calculation of the flux density (Giménez de Castro *et al.* 1999). SST has found that rapid pulsations are common to all sub-THz events observed (Kaufmann *et al.* 2001; Makhmutov *et al.* 2003; Raulin *et al.* 2003). Furthermore a new kind of burst emission was discovered exhibiting a THz component spectrally separated from the well known microwave component (Kaufmann *et al.*, 2004). The T-rays component was found for other events (Silva *et al.* 2007; Cristiani *et al.*, 2008; Kaufmann *et al.* 2009,). In other examples it appears as spectral frequency upturn trends during periods of certain events (Raulin *et al.* 2004; Lüthi, Ludi & Magun 2004).

The sub-THz time profiles and spectrum for the November 4, 2003 giant solar flare are shown in Figure 2. The spectrum shown at the inlet panel is for the middle major peak obtained with GHz data from Owens Valley Solar Array (OVSA) averaged over 30 s, and SST data. It shows another spectral component, peaking somewhere in the THz range, separated from the well known microwaves component. It is also observed that sub-second pulsations are superimposed onto the underlying flux time profile. A 20 s zoom near the middle major peak of the 40 ms resolution data at the bottom of Figure 2 shows pulses with significant amplitudes, similar at the two frequencies repeating at a rate close to 2 per second. The amplitude of the pulses relative to the mean flux ($\Delta$S/S) is of about 8% at 212 and 405 GHz, nearly all the time throughout the event duration. The pulses' amplitude are more than 10 times larger than the increase in system noise due to the burst contribution at the main peak i.e. of about 5 and 9 solar flux units (SFU), at 212 and 405 GHz, respectively (1 SFU = $10^{-22}$ W m$^{-2}$ Hz$^{-1}$).

Pulse rate count throughout the burst duration, was performed using a wavelet decomposition technique (Kurths & Schwarz 1994; Aschwanden *et al.* 1998; Giménez de Castro *et al.* 2001; Makhmutov *et al.* 2003; Raulin *et al.* 2003). For this analysis we use wavelets derived from a triangular Mother Wavelet which are representative of the pulses we want to observe (Bendjoya, Petit & Spahn 1993). The use of wavelets in solar flare analysis is not new (Krüger *et al.* 1994, Aschwanden *et al.* 1998, Giménez de Castro *et al.*, 2001, Makmuhtov *et al.* 2003). The method is useful to separate the original time series in different temporal scales. Every temporal scale is represented as a time series with 0 mean value, therefore, every significant local maximum can be considered a "pulse". As a requisite pulses are counted when their intensity exceeds a noise fluctuation threshold level set before the burst onset. The analysis was performed using 5 ms resolution data. We used temporal scales ranging from 40 to 160 ms and defined a criterion to avoid counting the same pulse in two different scales. The method was tested on time intervals with actually counted pulses, as illustrated in Figure 2(a). The results for the flux time profiles and for the pulse repetition rates, for the three events studied here, are shown in Figures 3, 4 and 5.

The identification of pulses superimposed onto the flux time profiles is strongly dependent on the quality of the atmospheric transmission. Transmission was extremely good for the November 4, 2003 solar burst observation, for which the time profiles were almost identical at 212 and 405 GHz both for fluxes and for the repetition rates (Figures 2(a) and 3). Atmospheric transmission was particularly bad for the November 2, 2003 observations, for which the pulsations were detectable only at 212 GHz (Figure 4). Poor transmission for the December 6, 2006 observations also limited the detection of pulsations at 405 GHz (Figure 5). Nevertheless it was possible to identify the presence of rapid pulsations in both events with qualitative amplitudes of about $\Delta S/S \approx$ 5-10%.

The results evidence a linear proportionality between the time profiles of pulse repetition rates with respect to the mean fluxes. The pulse repetition rate vs flux proportionalities are shown by the scatter diagrams of Figures 6, 7 and 8, for the events of November 4 and 2, 2003 and December 6, 2006, respectively. The data scattering is more pronounced for November 2, 2003 and December 6, 2006 when atmosphere transmission was poor. Therefore the suggested trend for clustering of points in groups might not be real, and cannot be discussed.

Data for the November 4, 2003 event (Figures 3 and 6) are particularly meaningful, because they were obtained with the best atmosphere transmission conditions at the site. They provide very large values for the linear correlation coefficients of flux vs pulse repetition rates. The linear fits to the diagrams suggest a convergence to zero flux for zero pulse/s. This trend may allow us to relate the solar burst sub-THz fluxes S related to the superimposed pulse rates, R, as

$$S = k \, R \qquad\qquad\qquad (1)$$

where k has the dimension of energy (density per frequency band). The scatter diagrams shown in Figures 6, 7, and 8 suggest that k ≈ 4000 SFU . s (i.e., 4 $10^{-19}$ J $m^{-2}$ $Hz^{-1}$) at 405 GHz, and k ≈ 800-1200 SFU . s (i.e., 0.8-1.2 $10^{-19}$ J $m^{-2}$ $Hz^{-1}$) at 212 GHz.

## 3. Distinctions of rapid pulse features at sub-THz and microwaves

The three events analyzed here had no microwave counterparts available for comparison with sufficiently high sensitivity and time resolution. However, for the November 4, 2003 burst, despite of saturation during the bulk of emission, the onset and decay phases were well observed at 44 GHz by the Itapetinga radio-telescope (Brazil), with high sensitivity and time resolution. Figure 9 shows the 100 ms flux time profiles at 44 GHz (top) and the repetition pulse rate time profile (bottom).

Detailed comparison of sub-THz and GHz pulse time profiles are shown for the burst onset in Figure 10. The flux scale is the same for the three frequencies to illustrate that it is not possible to make a meaningful one-to-one pulse comparison in the two frequency ranges because the sub-THz pulse intensities are close to their smaller limiting values and about 20 times weaker compared to the GHz intensities. The 44 GHz pulse amplitudes relative to the mean flux $\Delta S/S$ are about 80%, nearly ten times larger than $\Delta S/S$ at the two sub-THz frequencies. It compares well to large $\Delta S/S$ values obtained for other bursts at 48 GHz (Raulin *et al.* 1998). Comparing Figure 10 to Figure 3 we note that the onset of the 44 GHz pulses begins about two minutes earlier than that of the sub-THz, which can be a consequence of the much larger amplitude of the GHz pulses. The pulse rates at 18 43 UT and 18 49 UT are of about 0.5-1.0 per second, comparable for the sub-THz and GHz frequencies during the periods of simultaneous observation.

The flux vs pulse rates relationship, S ∝ k R, known at microwaves, shows a consistent qualitative trend on a widely scattered data points obtained for various events (Kaufmann *et al.* 1980; Qin *et al.* 1996; Huang, Qin & Yao 1996; Raulin *et al.* 1998). It has been known that the relative pulse amplitudes $\Delta S/S$ observed at discrete microwave frequencies range from few percent at about 10 GHz up to 50% at 90 GHz, shown in Figure 11 (plotted after Correia & Kaufmann 1987, with new data added). This might provide a consistent explanation for the large pulse amplitude observed at 44 GHz on November 4, 2003 burst. Unfortunately there were no high time resolution data available at any other microwave frequencies. On the other hand, there is a clear distinction in the sub-THz relative pulse amplitudes which are considerably smaller than the extrapolations (Correia & Kaufmann 1987), as well as in comparison to the 44 GHz pulses observed before and after saturation.

These results suggest that, although there might exist a causal relationship between superimposed pulses observed at microwaves and at sub-THz bursts, they are probably produced at different physical locations and/or by different emission mechanisms.

## 4. Discussion

The pulses superimposed onto the broadband solar bursts emission might be interpreted as a response to physical processes at the origin of the flare. Different mechanisms may be assigned for emissions producing the distinct spectral structures (Figure 1). Maximum microwave fluxes are at frequencies in the range 5-20 GHz (Guidice

& Castelli 1975), and some times extending up to 100 GHz (Croom 1971, Nakajima *et al.* 1985; Bastian, Fleishman & Gary 2007). The spectral component B in Figure 1 is usually attributed to electron gyrosynchrotron radiation (Dulk 1985; Bastian, Benz & Gary 1998). At meter-decimeter wavelengths the burst emission intensity increases for smaller frequencies (spectral structure A in Figure 1). This component is usually attributed to plasma emission excited by fast electrons or by shock waves (Wild & Smerd 1972). The nature of the THz burst emission (structure C in Figure 1) is not well understood. It might be attributed to a distinct synchrotron emission by ultrarelativistic electrons (Kaufmann *et al.* 2004; 2009; Kaufmann & Raulin 2006). Often the burst emission at the U-shaped minima (roughly 0.5-5 GHz, see Figure 1) may contain contributions of different origin added together (Alaart *et al.* 1990; Altyntsev *et al.* 2003). Mixed contributions from different mechanisms may also occur at the 30-200 GHz spectral minimum.

Quasi periodic pulsations of gyrosynchrotron emission at microwaves are known to cover a wide range of time scales - from hundreds of milliseconds to tens of seconds - and are interpreted in terms of modulation of source parameters undergoing MHD oscillations and/or modulation of electron acceleration or injection (Fleishman, Bastian & Gary 2008). The proportionality between flux and pulse rate at millimeter waves (Kaufmann *et al.* 1980; Raulin *et al.* 1998) supports the hypothesis that impulsive bursts might be a response to rapid injections quasi-quantized in energy (Kaufmann *et al.* 1980, Sturrock *et al.* 1984). The ripple structure superimposed onto impulsive bursts time profiles might be the result of the convolution of pulses of a given lifetime with a variable repetition rate (Loran *et al.* 1985).

The flux vs pulse rate correlation was also suggested to be a response to the inverse correlation between the saturation time of a wave-particle plasma instability originating the burst and the radiation time (Huang, Qin & Yao 1996). This effect might also be attributed to the modulation of flare kernel emission by magnetic field oscillations driven by Alfvén waves, whose period decreases for larger temperatures (Qin *et al.* 1996). This interpretation assumes that the GHz emission is produced by gyro-synchrotron radiation from electrons moving in the active region magnetic field. However, Fleishman, Bastian & Gary (2008) interpretation of a pulsating microwave burst, based on Dulk (1985) gyro-synchrotron emission formulas, has shown that for sausage mode MHD oscillations the flux variations in the optically thick part of the spectrum (i.e., frequencies smaller than the spectral peak in Figure 1 structure B) should be anti-correlated in time with variations observed in the optically thin part of the spectrum (i.e. for frequencies higher that the spectral peak). This effect was not observed, leading them to suggest that the pulsations were associated to quasi-periodic injection of emitting electrons.

It has been found qualitatively that the relative pulse amplitude $\Delta S/S$ increase for higher frequencies, as shown in Figure 11 (Kaufmann *et al.* 1985; Correia & Kaufmann 1987). This trend agree with the large relative pulse amplitudes observed at 44 GHz (about 80%) at the onset and decay of the November 4, 2003 event (see Figures 9 and 10) nearly ten times larger than the sub-THz pulses relative amplitudes. However the observed sub-THz $\Delta S/S$ for the three events analyzed here are considerably smaller than expected when extrapolated from GHz observations (Kaufmann *et al.* 1985; Correia & Kaufmann 1987) (Figure 11).

The sub-THz component might be described as the optically thick part of the synchrotron radiation emitted by highly energetic electrons (> 10 MeV) moving in a strong magnetic field (≥ 1000 Gauss) (Kaufmann *et al.* 2004; 2009; Kaufmann & Raulin 2006; Silva *et al.* 2007; Trottet *et al.* 2008, Cristiani *et al*, 2008). It should be somehow causally related but physically distinct from the GHz component. The linear relationship between sub-THz fluxes and pulse repetition rates might be a direct indication that the flare accelerator injects discrete quantized energy packets. The emissions at microwaves as well as at other ranges of energy might present proportional relations to these repetitive energetic injections that need more investigation.

## 5. Concluding remarks

It has been shown that rapid pulsations are found superimposed onto all events that present the THz spectral component. Pulse rates estimates were performed using wavelet decomposition at two sub-THz frequencies, 212 and 405 GHz. The pulse count criteria were set by fluctuation thresholds determined before the events. Pulses were particularly well defined for one event (November 4, 2003). For the other two events, the pulse identification was limited by the poor quality of atmosphere transmission, affecting mainly the 405 GHz data.

Similar small $\Delta S/S$ were found for two other solar flares observed at sub-THz frequencies, on April 6, 2001 (Kaufmann *et al.* 2002), and on August 25, 2001 (Raulin *et al.* 2003). They were not included in this study because there were indications that the sub-THz emissions might contain contributions from both GHz and THz spectral components (B and C in Figure 1).

Nevertheless the analysis revealed two relevant properties of the sub-THz pulses: (a) the mean fluxes (S) are proportional to the pulse repetition rates (R) according to the simple relationship $S \approx k$ R; (b) the pulse amplitudes relative to the mean flux, $\Delta S/S$ are small, ranging 5-10% throughout the bursts durations. It follows that the flux vs pulse rate relationship might be interpreted as response to discrete and successive energetic injections, quantized in energy: $4\ 10^{-19}$ J m$^{-2}$ Hz$^{-1}$ at 405 GHz, and 0.8-1.2 $10^{-19}$ J m$^{-2}$ Hz$^{-1}$ at 212 GHz.

The result (a) confirms the flux vs pulse rate qualitative relationship known in GHz bursts (Kaufmann *et al.* 1980; Qin *et al.* 1996; Huang, Qin & Yao 1996; Raulin *et al.* 1998), and shows a good fit to a linear correlation. The correlation coefficients worsen when atmospheric transmission is poor, especially at 405 GHz. The sub-THz flux increase with frequency might be associated to the optically thick part of synchrotron emission by high energy electrons (> 10 MeV).

In the GHz range, however, the relative pulse amplitude ($\Delta S/S$) increases substantially with frequency (Correia & Kaufmann 1987, Huang, Qin & Yao 1996, see Figure 11), an effect that was confirmed by observations at 44 GHz for the onset and late decay of the November 4, 2003 burst (Figures 9 and 10). For the sub-THz events, however, the observed $\Delta S/S$ are comparable at the two frequencies (212 and 405 GHz) and considerably smaller than expected from extrapolated values. This apparent qualitative contradiction might be an indication that two distinct emission mechanisms produce the spectrally separated GHz and THz components.

The production of the two GHz and THz spectral components (components B and C in Figure 1) might be explained by conceiving independent accelerators at the flaring source, producing different energy electrons nearly simultaneously, emitting the two spectral components. The spectral components B and C in Figure 1 would be representative of two synchrotron emission spectra, by electrons with mildly relativistic energies and ultrarelativistic energies, respectively. However the observations available are still too limited to support this possibility. Another scenario suggested by Wild & Smerd (1972) might be adopted, placing a single accelerator closer to a single polarity foot-point injecting electrons into a magnetic morphology split into two separate loops, one low altitude with stronger field, another weaker field, higher above the solar surface, originating the two synchrotron spectral components, on peaking in the GHz and another in the THz range. Although this possibility was suggested in the discussion of the November 2, 2003 event (Silva *et al.* 2007), it requires very strong magnetic fields, of the order of the largest ones observed to date. In both scenarios the relative pulse amplitude (ΔS/S) features in the GHz and sub-THz ranges remain to be explained. An alternate interpretation take into account the contribution to spectral emissions from electron beam instabilities observed in laboratory accelerators (Nodvick & Saxon 1954, Williams 2002; Carr *et al.* 2002; Byrd *et al.* 2002) proposed to happen in solar flares (Kaufmann & Raulin 2006; Klopf 2008).

Observations are needed at higher frequencies in the THz range, as well as with higher time resolution and sensitivity in the GHz range. High sensitivity and time resolution solar patrol polarimeters at 45 and 90 GHz are planned for El Leoncito Observatory with the support of the Brazilian research agency FAPESP and Argentinean agency CONICET. Solar photometry at two THz frequencies are considered by the experiment DESIR (Detection of Solar eruptive Infrared Radiation) on the French-China satellite SMESE (SMall Explorer for the study of Solar Eruptions) (Vial *et al.* 2008). These new experiments are expected to bring new clues to understand the origin, the nature, and the relationship between flare emission at GHz, THz, and other ranges of energy, in response to repetitive discrete energetic injections, and their possible association with analogous physical processes in laboratory accelerators.

*Acknowledgements*. We gratefully acknowledge the remarks given by one anonymous referee helping the improvement of this paper presentation. These researches were partially supported by the Brazilian agencies FAPESP, CNPq, Mackpesquisa, and Argentina agency CONICET.

**Captions to the figures**

**Figure 1** – Schematic representation of solar radio burst emissions, from metric to submillimetric wavelengths, adapted from a plot given by Castelli (1972) for U-shaped spectra (up to 30 GHz). The newly discovered THz component (Kaufmann *et al.* 2004) gives a more general W-shaped radio emission spectra build up from three different emission mechanisms: A, B and C.

**Figure 2** – The November 4, 2003 solar burst. Top panel shows the time profiles obtained by SST at sub-THz frequencies. The inlet shows the spectral trend for middle temporal peak, evidencing two spectrally separated components, at microwaves and at sub-THz. Bottom panel exhibits a 20 s zoom of 40 ms data near the middle maximum structure showing the rapid superimposed pulsations.

**Figure 3** – 4 November 2003 flare. (a) 212 GHz 40 ms flux time profile (top), and pulse rate profile (bottom). (b) 405 GHz 40 ms flux time profile (top), and pulse rate profile (bottom). Pulse rate obtained on 5 ms data, every 5 s interval, counted using wavelet decomposition.

**Figure 4** - 2 November 2003 flare. The sub-THz data were considerably prejudiced by poor atmosphere transmission. Pulses at 405 GHz were below the noise fluctuation threshold set prior to the event, and could not be counted. The start time was not well defined. We show here only the 212 GHz 40 ms flux time profile (top), and pulse rate profile (bottom). Pulse rate obtained on 5 ms data, every 5 s interval, counted using wavelet decomposition.

**Figure 5** - 6 December 2006 flare. (a) 212 GHz 40 ms flux time profile (top), and pulse rate profile (bottom). (b) 405 GHz 40 ms flux time profile (top), and pulse rate profile (bottom). Pulse rate obtained on 5 ms data, every 5 s interval, counted using wavelet decomposition. Due to relatively large atmospheric attenuation, most of the pulses at 405 GHz were below the noise fluctuation threshold set prior to the event. They could be counted only partially in the main impulsive phase.

**Figure 6** – Scatter diagram showing the correlation between fluxes and pulse repetition rates for the 4 November 2003 event, at 212 GHz (diamonds) and 405 GHz (circles). The best fit linear correlation coefficients are of 0.975 and 0.953 at 212 and 405 GHz respectively.

**Figure 7** - Scatter diagram showing the correlation between fluxes and pulse repetition rates for the 2 November 2003 event, at 212 GHz. Despite of the data scattering caused by the poor atmosphere transmission, the best fit linear correlation coefficients is of 0.74 at 212 GHz.

**Figure 8** – Scatter diagram showing the correlation between fluxes and pulse repetition rates for the 6 December 2006 event, at 212 GHz (diamonds) and 405 GHz (circles). Data scattering is due to fluctuation in poor atmospheric transmission. The best fit linear correlation coefficient is of 0.87 at 212 GHz. The 405 GHz data points are too scattered to be meaningful (correlation coefficient of 0.61), but they do indicate a qualitative trend, similar to the result shown in Figure 6.

**Figure 9 -** 4 November 2003 flare onset and decay phases at 44 GHz 100 ms flux time profile (top), and pulse rate profile (bottom). Pulse rate were obtained every 5 s interval, counted using wavelet decomposition.

**Figure 10** – Expanded view of 4 November 2003 time profiles at the burst onset, just previous to the saturation at 44 GHz, illustrating how stronger were the GHz pulses as compared to the sub-THz pulses.

**Figure 11** – Qualitative scatter diagram showing the relative pulse amplitude ($\Delta S/S$) dependence in frequency, after Correia and Kaufmann (1987) (●); to which have been added data derived from Qin *et al.* (1996) (□); Huang, Qin & Yao (1996) (O); Raulin *et al.* (1998) (⊕); and from Nakajima (2000) (∇). The 4 November 2003 data analyzed here are shown by the symbol (∗).

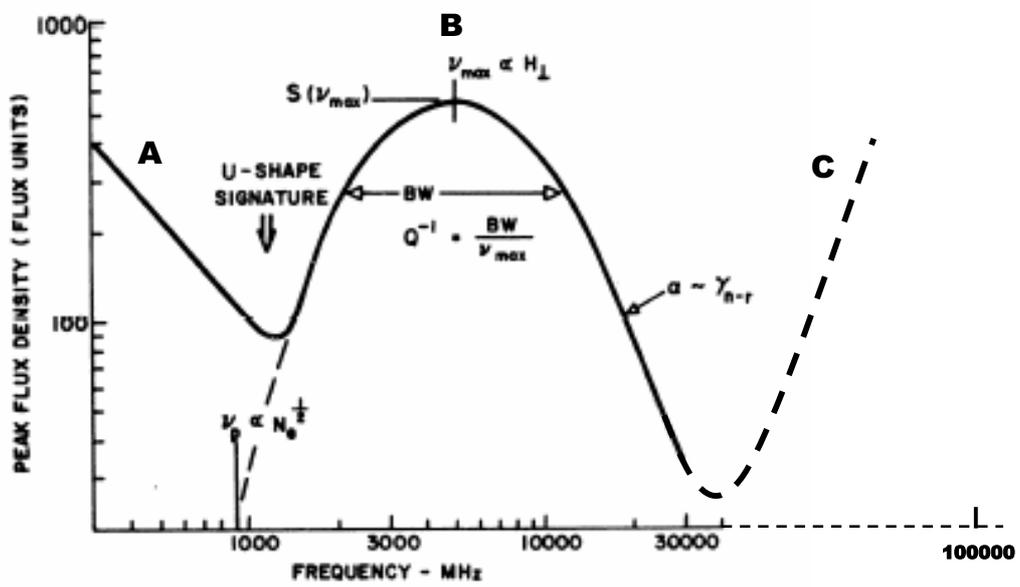

FIG. 1

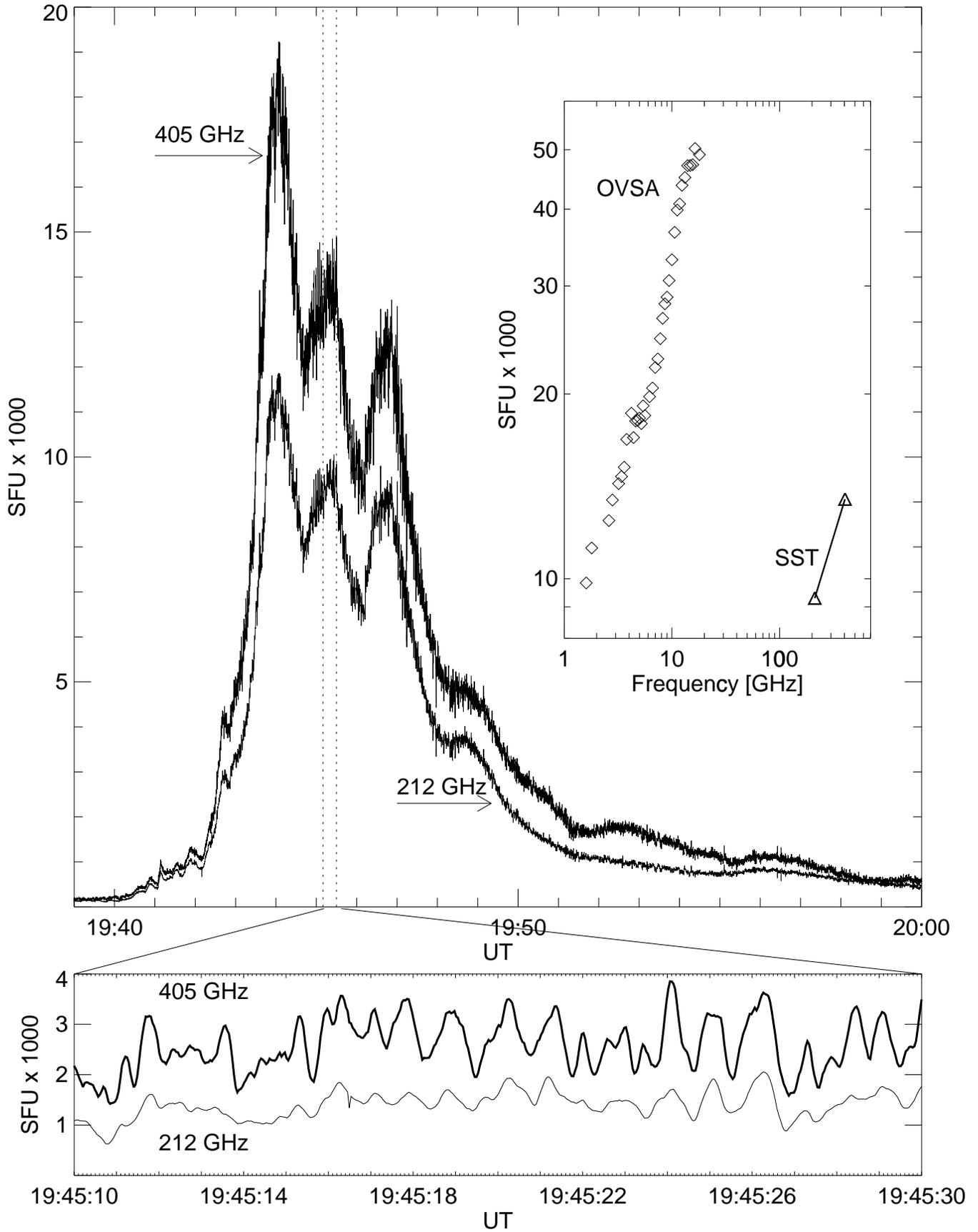

FIG. 2

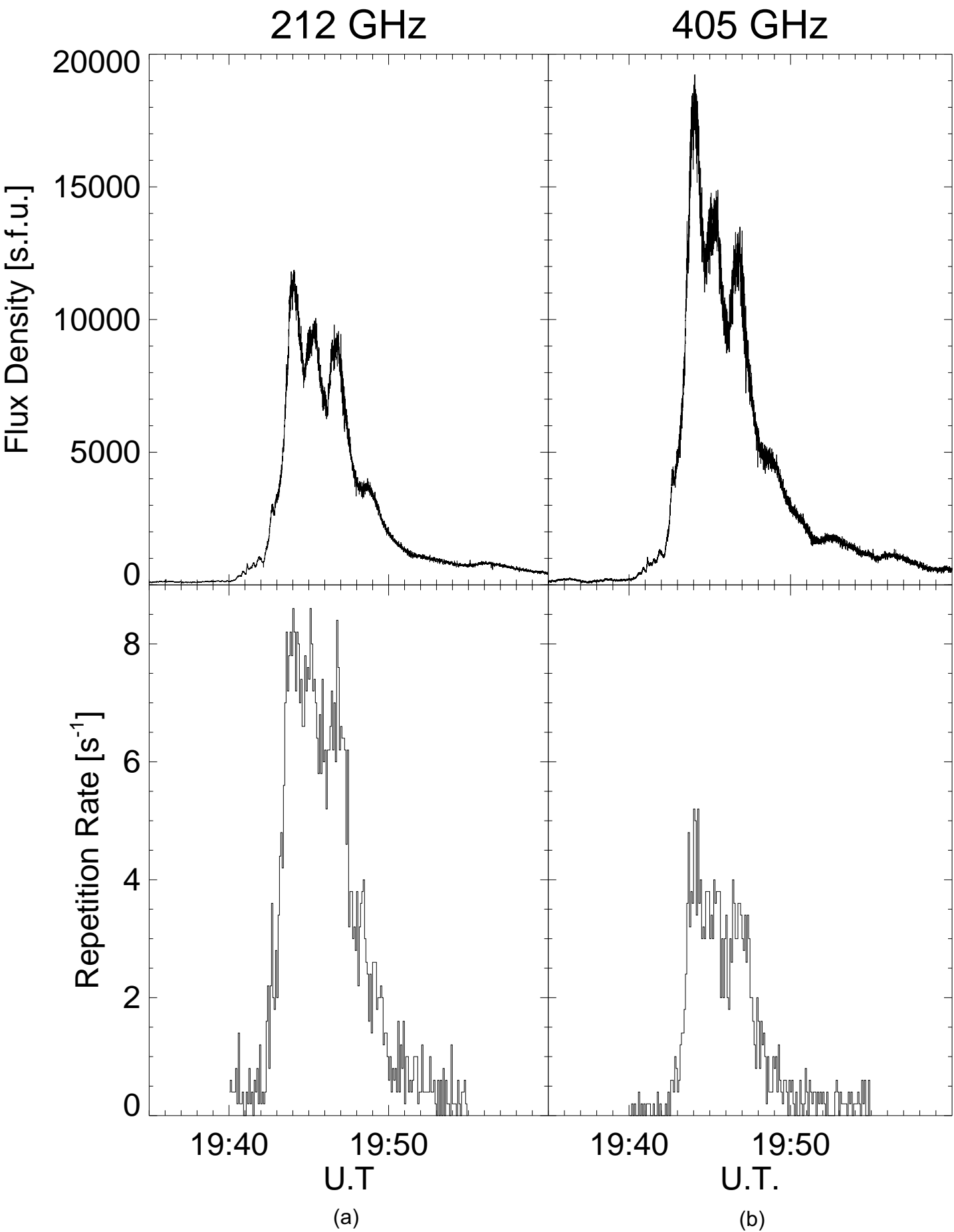

FIG. 3

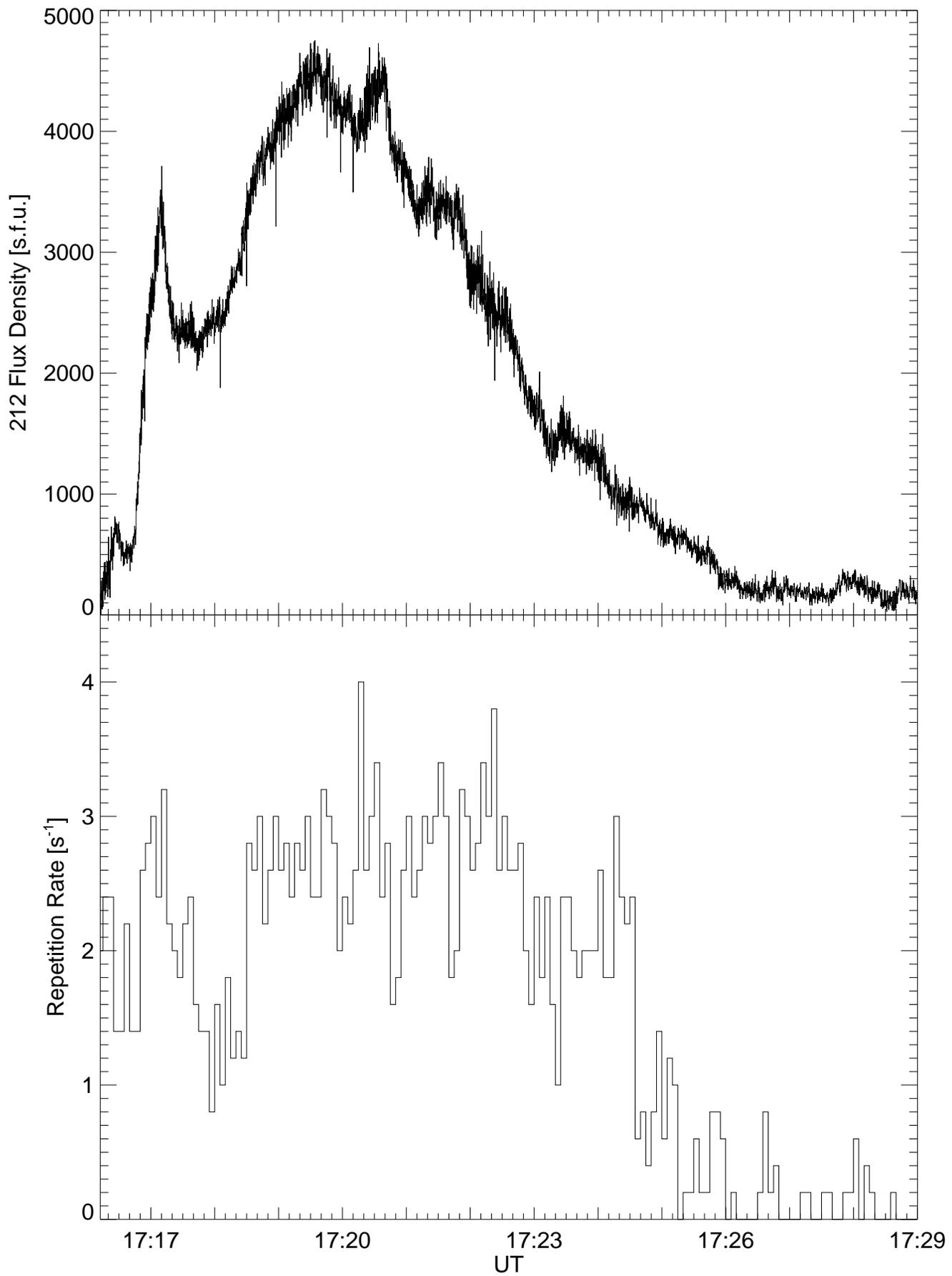

FIG. 4

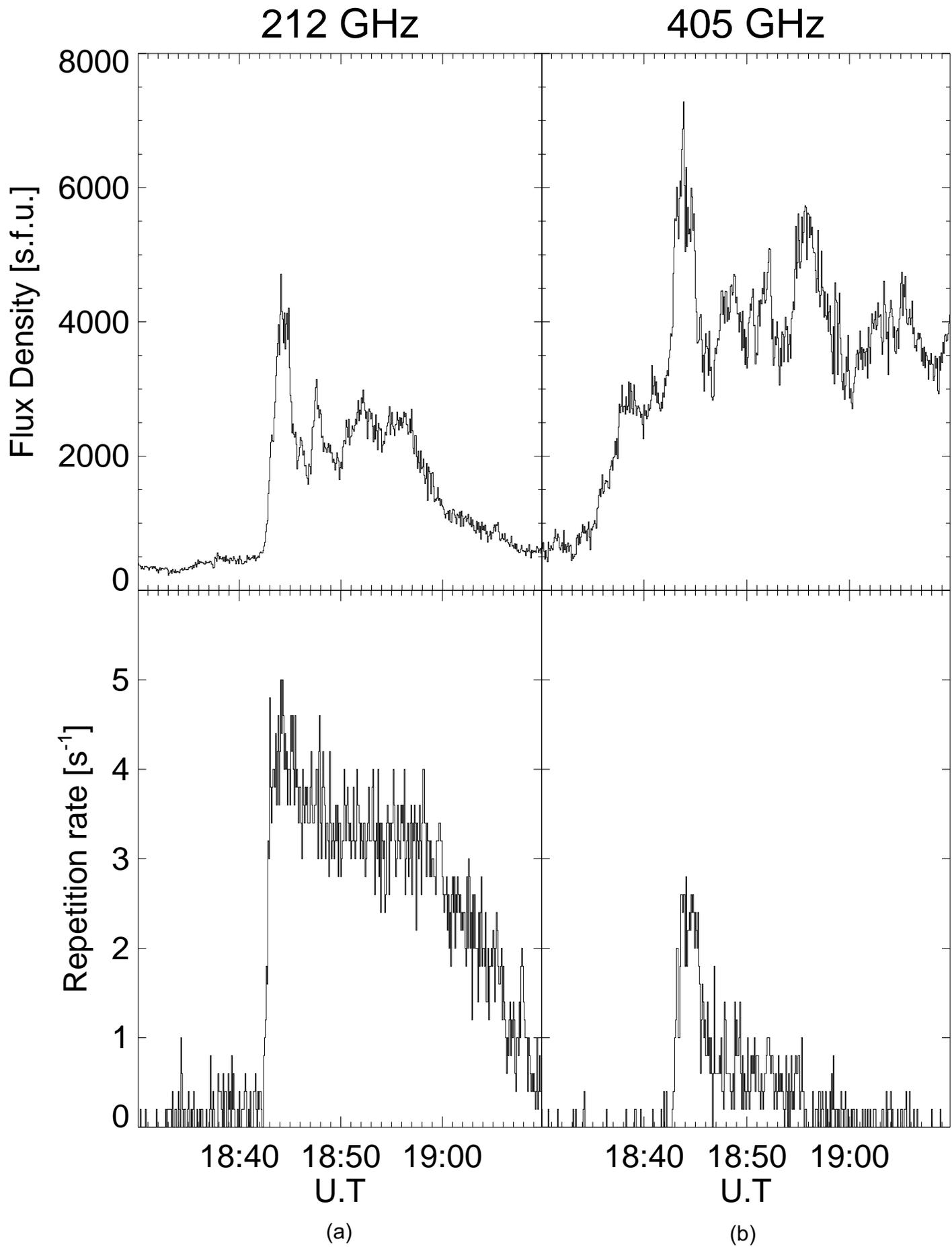

FIG. 5

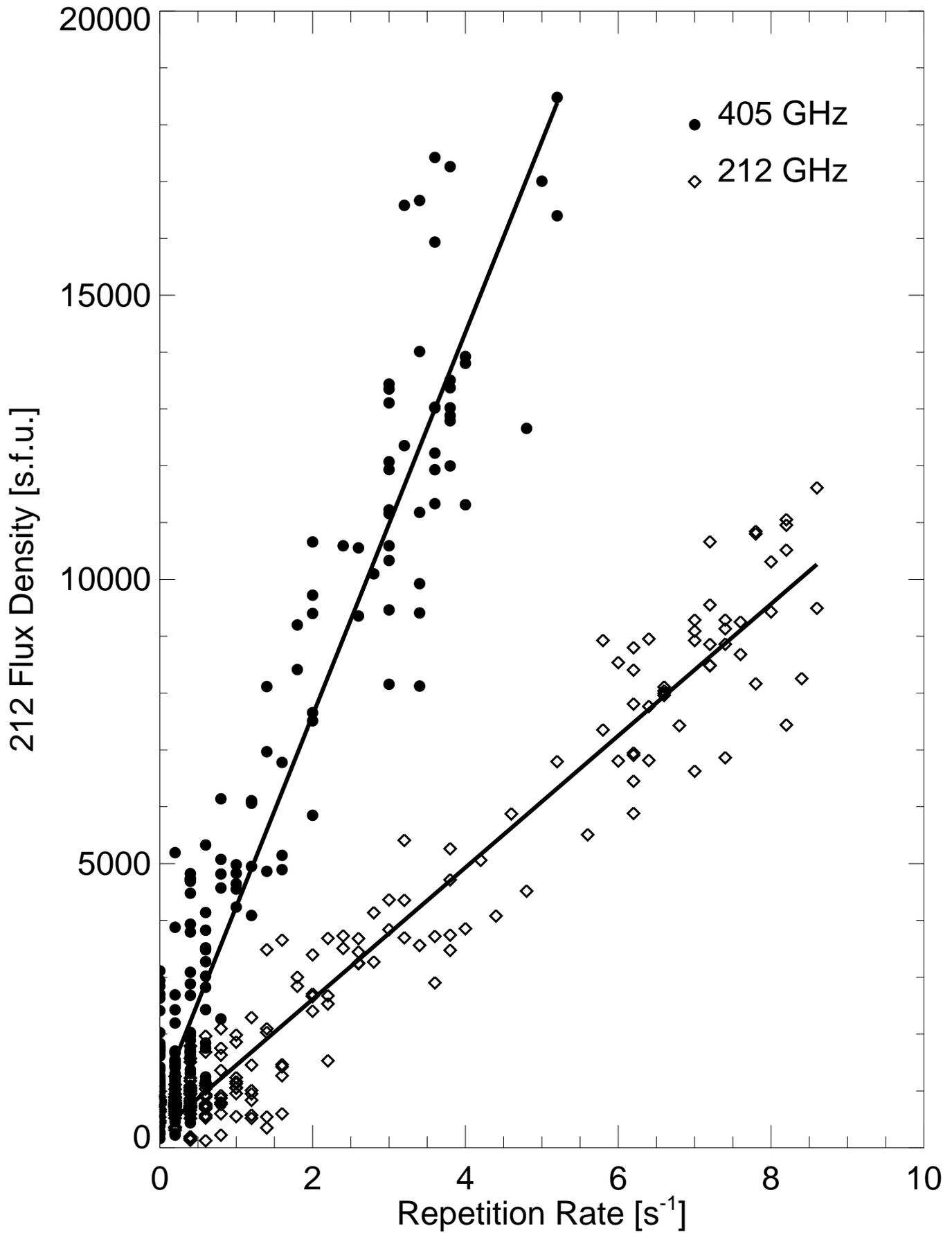

FIG. 6

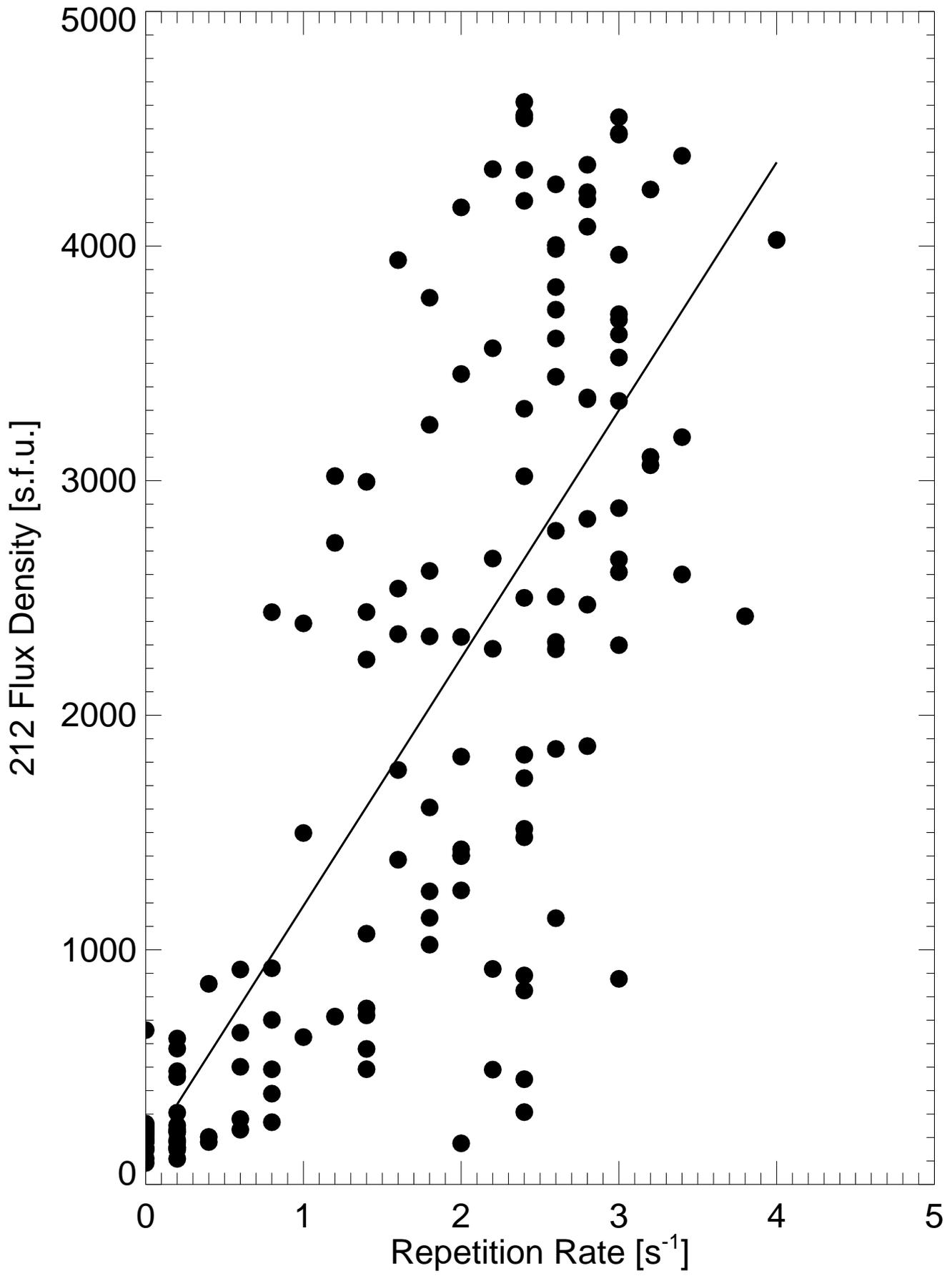

FIG. 7

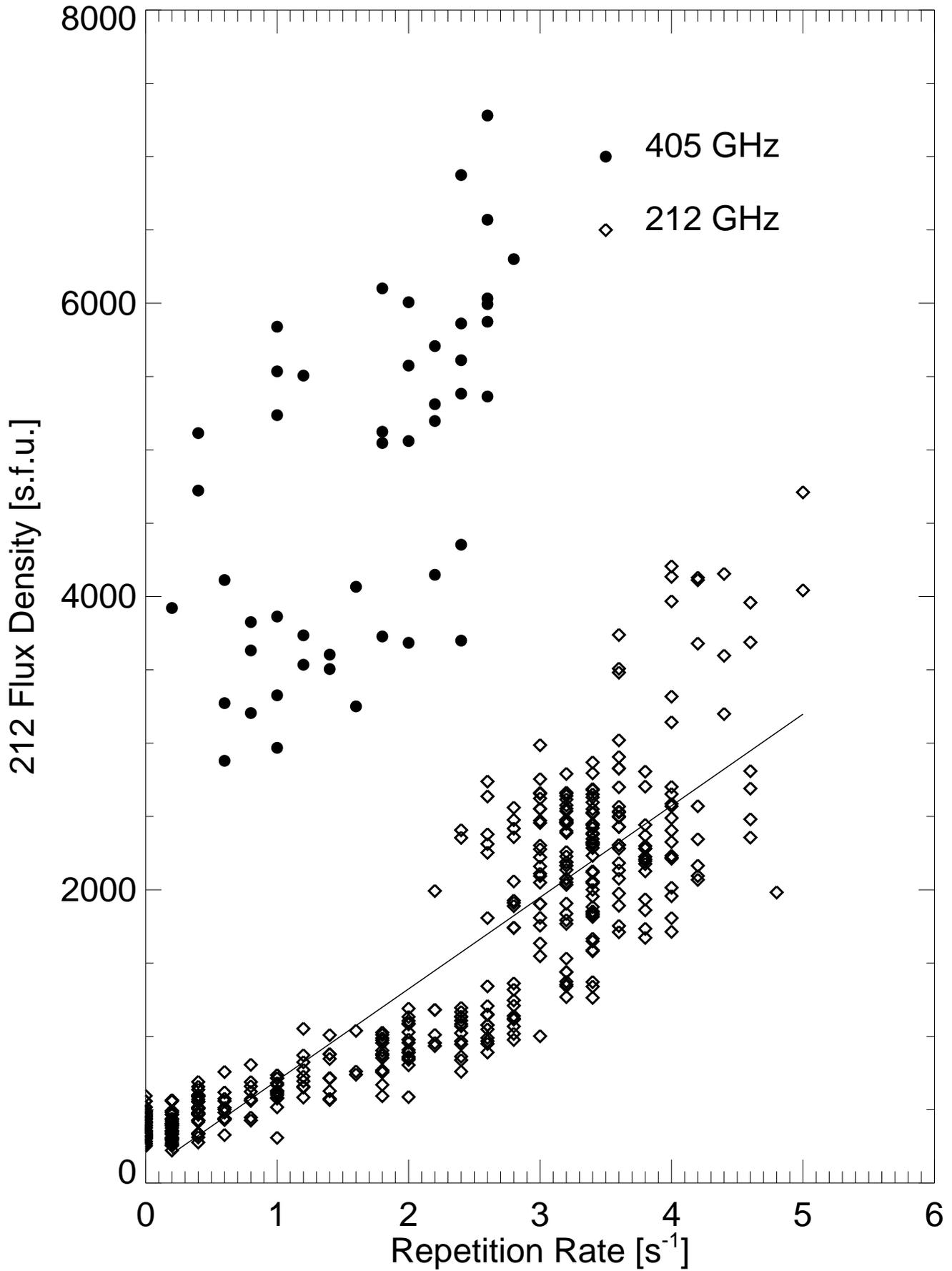

FIG. 8

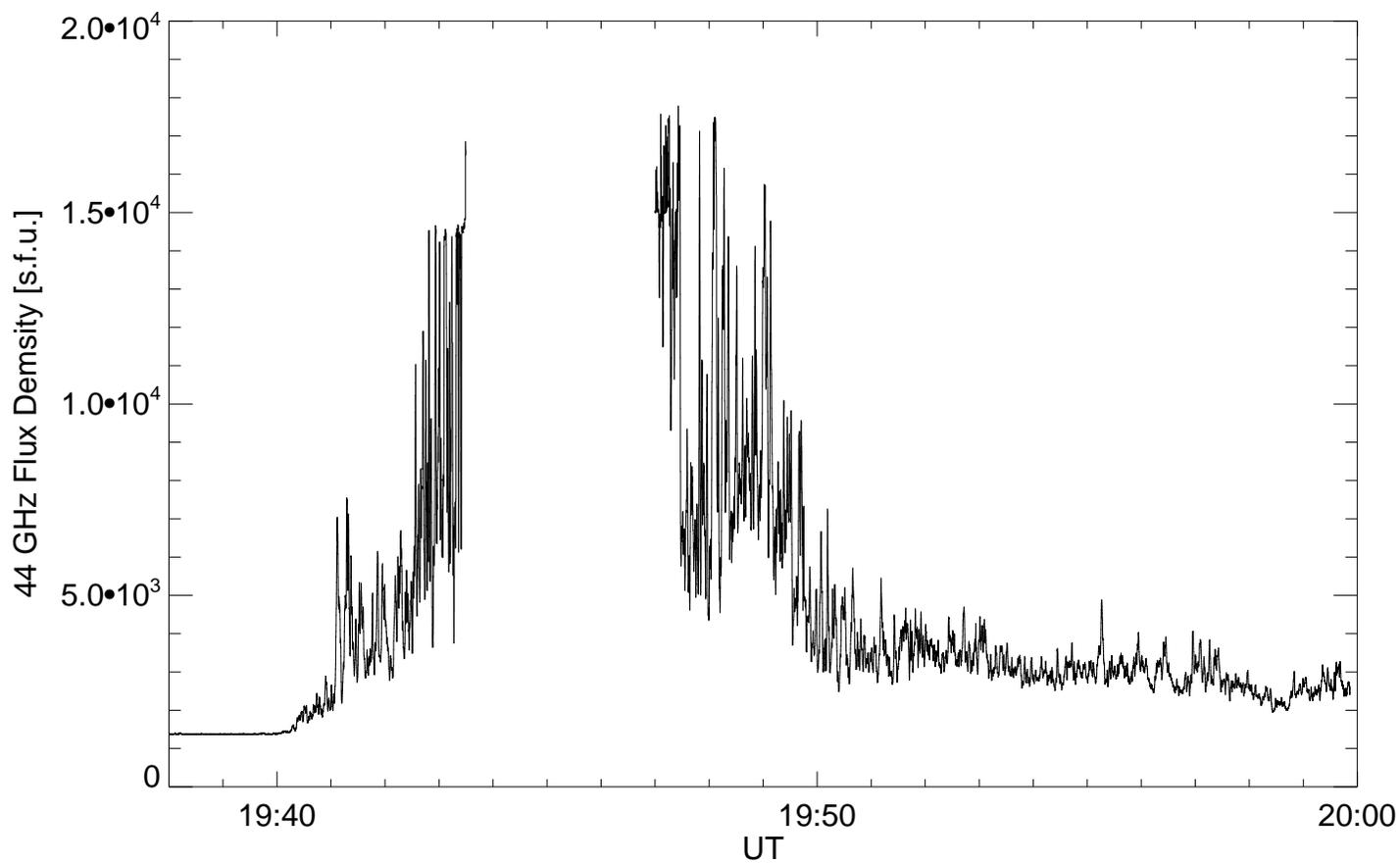

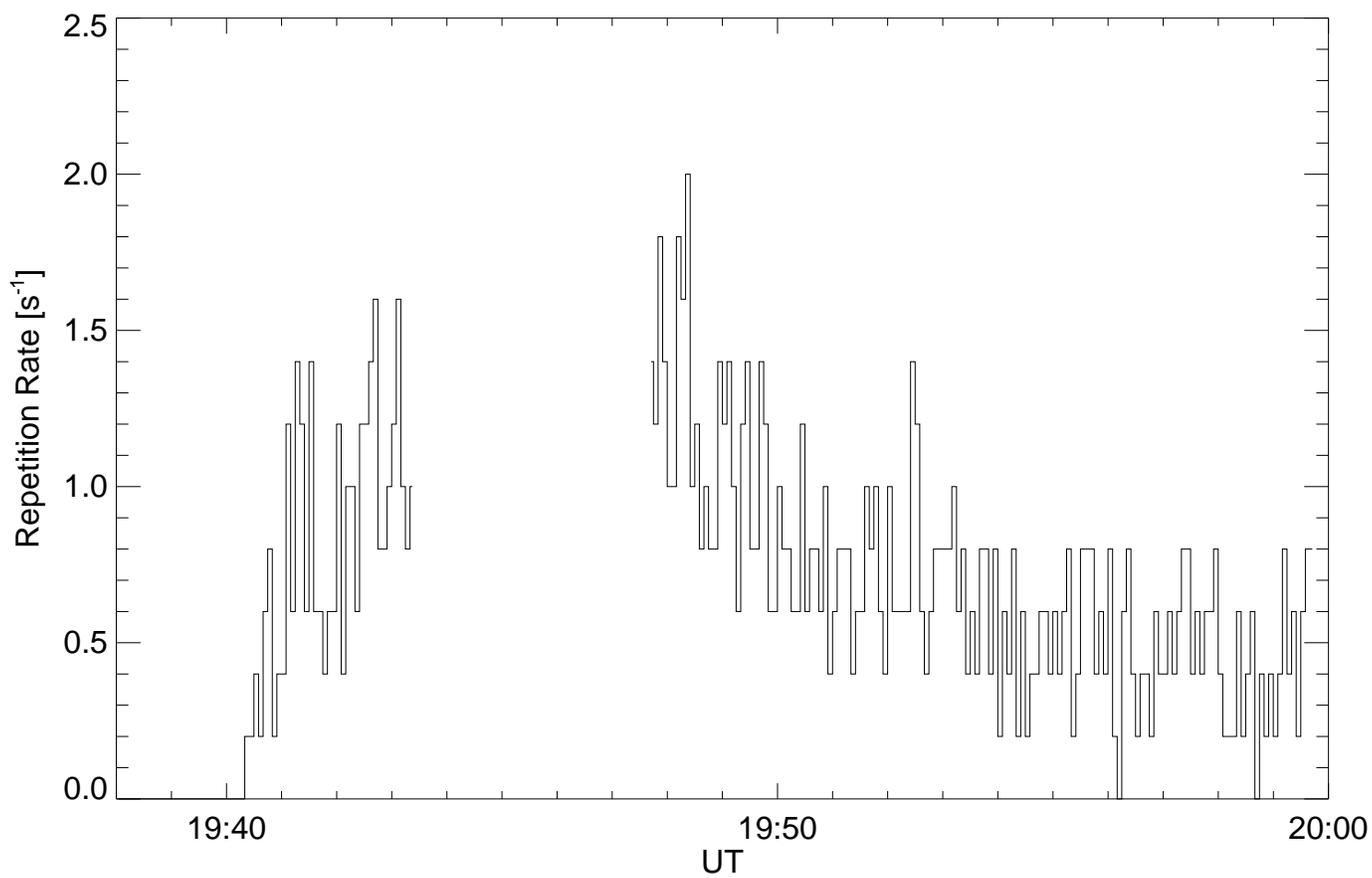

FIG. 9

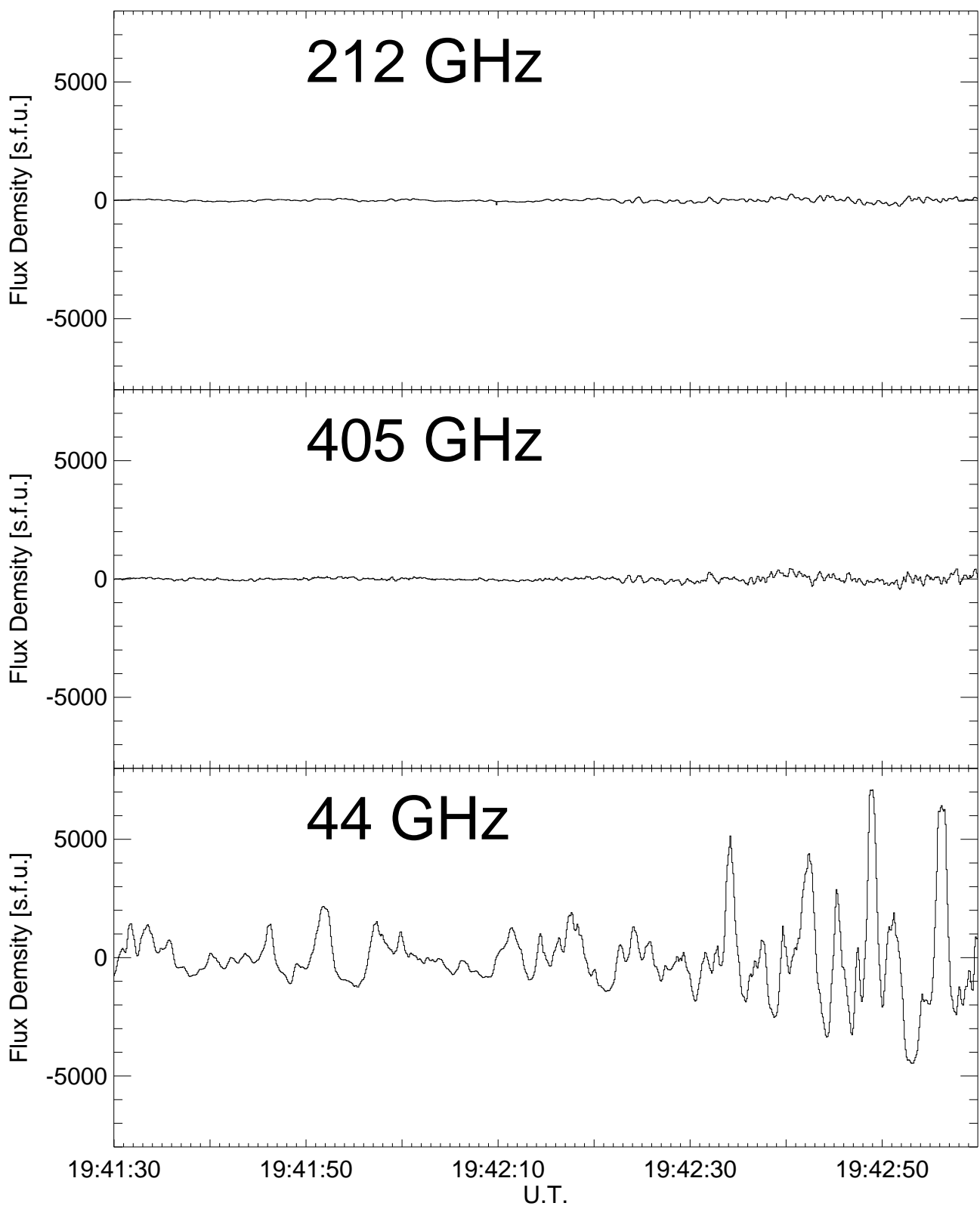

FIG. 10

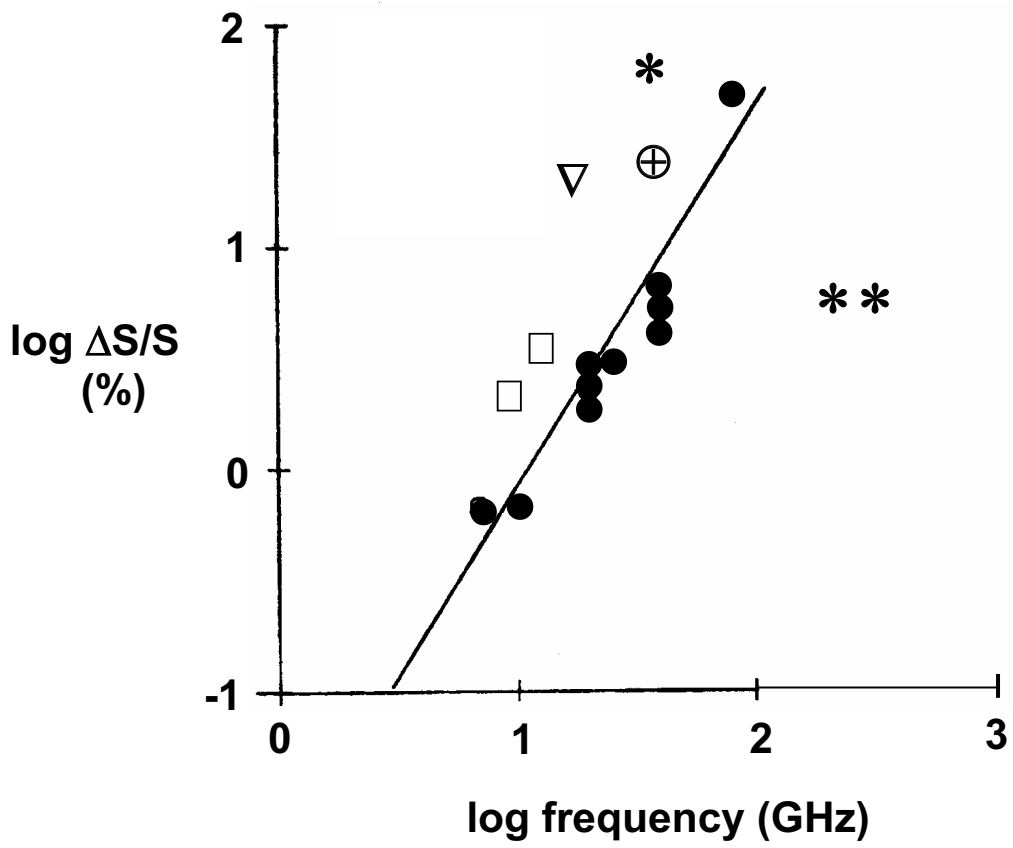

FIG. 11